\documentclass[prl,twocolumn, showpacs]{revtex4}%twocolumn,
\usepackage{enumerate}

\usepackage{graphicx}

\usepackage{color}

\begin{document}

\definecolor{Faina}{rgb}{0.4,0.5,0.9}
\definecolor{Gold}{rgb}{0.5,0.23,0.25}
%\textcolor{blue}{{\it uniform}}
%
\title{Semigroup evolution in Wigner Weisskopf pole approximation
with Markovian spectral coupling
}
\author{
F. Shikerman~$^{(1)}$, A. Peer~$^{(1)}$, L. P.
Horwitz~$^{(1,2,3,4)}$
\\ {\it Department of Physics, Bar-Ilan University,
Israel, Ramat-Gan, 52900$~^{(1)}$}\\ {\it School of Physics,
Tel-Aviv University, Israel, Ramat-Aviv, 69978$~^{(2)}$}\\ {\it
Department of Physics, Ariel University Center of Samaria, Israel,
Ariel, 40700$~^{(3)}$} \\{\it IYAR, Israel Institute for Advanced
Research, Rehovot, Israel$~^{(4)}$}}

\begin{abstract}
We establish the relation between the Wigner-Weisskopf theory for
the description of an unstable system and the theory of coupling to
an environment. According to the Wigner-Weisskopf general approach,
even within the pole approximation the evolution of a total system
subspace is not an exact semigroup for multichannel decay, unless
the projectors into eigenstates of the reduced evolution generator
$W(z)$ are orthogonal. With multichannel decay, the projectors must
be evaluated at different pole locations $z_\alpha\neq z_\beta$, and
since the orthogonality relation does not generally hold at
different values of z, the semigroup evolution is a poor
approximation for the multi-channel decay, even for very weak
coupling. Nevertheless, if the theory is generalized to take into
account interactions with an environment, one can ensure
orthogonality of the $W(z)$ projectors regardless the number of the
poles. Such a possibility occurs when $W(z)$, and hence its
eigenvectors, are independent of z, which corresponds to the
Markovian limit of the coupling to the continuum spectrum.

\end{abstract}
\maketitle
%\tableofcontents

\section{Introduction}

 Many physical systems
demonstrate instability, i.e., a transition from a relatively stable
state to a final state which in general corresponds to a system with
many identifiable degrees of freedom. Such occurs, for example, in
particle decay or radiative atomic transitions. In many cases the
process observed has the semigroup property, i.e., the operators
generating the evolution on the Hilbert space of quantum states
satisfy the composition law
\begin{equation}
Z(t_1)Z(t_2) = Z(t_1 + t_2)\label{Eq1}\end{equation}
for $t_1,t_2\geq 0$, and for which the operators $Z(t)$ do not have
an inverse and are decreasing.  This relation is a general form of
the well-known exponential decay law, for which Gamow \cite{Gamow}
constructed a phenomenological Schr\"{o}dinger equation with complex
energy eigenvalue of negative imaginary part. Weisskopf and Wigner
\cite{WW} formulated a basic theory which approximately reproduced
the Gamow result in second order perturbation for a single channel
decay. In their original formulation, the survival amplitude of a
quantum state $|\psi\rangle$ is given as
\begin{equation}
U^{red}(t) = \langle\psi| e^{-iHt}|\psi\rangle,\label{Eq2}
\end{equation}
where $H$ is the full Hamiltonian of a system consisting of an
unperturbed part $H_0$ for which $|\psi\rangle$ is an eigenstate,
and a perturbation $V$ is understood to induce a transition to an
infinite number of final states with a continuous spectrum. The
Laplace transform of Eq.~(\ref{Eq2}) provides an expression
corresponding to the Green's function (or resolvent kernel) for the
Schr\"{o}dinger evolution. In a single channel decay problem the
pole approximation for the inverse Laplace transform results in an
approximate semigroup property for $U^{red}(t)$, which is equivalent
to the perturbative analysis of Weisskopf and Wigner. However, in
case of the two or more channel decay, such as the neutral K meson
decay (for which there is CP symnmetry breaking), it has been shown
that the pole approximation does not reproduce the semigroup
evolution observed in \cite{Fermilab}. To account for semigroup
behavior application has been made of a quantum form \cite{SHE,RSHP}
of the classical Lax-Phillips theory \cite{R} (see also \cite{SH}
and references therein) and its generalizations \cite{Yossi}. This
theory achieves an exact semigroup law by imbedding the usual
quantum theory in a larger Hilbert space, consisting of a direct
integral of a family of Hilbert spaces of usual type, foliated
according to the time parameter.
%Since the subspaces of this space are defined according to
%boundaries in time as well as space, the generator of motion may
%become non-selfadjoint and acquire complex eigenvalues \cite{RSHP}.
%The structure is characterized by invariant incoming and outgoing
%subspaces, and resonant states are represented in the complement of
%these subspaces. Originally developed by Lax and Phillips \cite{R}
%to study classical wave equations, Strauss et al \cite{SHE}
%succeeded to extend its application to the quantum theory.
The result was particularly effective and straightforward for
treating quantum mechanical systems with Hamiltonians of unbounded
spectrum \cite{R1,R2}. Its generalization for problems with
semibounded spectrum \cite{Yossi} made it clear that the imbedding
associated with the Lax-Phillips theory effectively introduces many
additional degrees of freedom. In this paper, we show that semigroup
evolution may be obtained working in the framework of the usual
quantum theory through an explicit coupling to many environmental
degrees of freedom. Thus, we provide a physical framework accounting
for the
mathematical structure of the Lax-Phillips theory. %In this paper, we
%work in the framework of the usual quantum theory, and show that the
%semigroup behavior can be achieved through an explicit coupling to
%the many degrees of freedom carried by an environment. We have
%therefore provided a physical framework accounting for the
%mathematical structure of the Lax-Phillips theory, a framework also
%entering in the dilation formalism developed by Maasen\cite{Maasen}
%in his constructive realization in a Fock space of the
%Nagy-Foias-Kolmogorov \cite{R3} embedding developed by Lax and
%Phillips. In particular, we show in this paper that the original
%formulation of Weisskopf and Wigner can result in a semigroup law if
%the transition amplitudes entering in the computations can be
%understood in terms of realistic interactions with an environment.\\

We study the Wigner-Weisskopf pole approximation theory \cite{WW} in
the framework of the Lee-Friedrichs model \cite{Lee}. This model
consists of a subspace of discrete states interacting with a
continuum subspace of states, for which there is no direct
continuum-continuum interaction. It is convenient to take the
discrete-discrete interaction to vanish as well. The association of
the continuum with an environment constitutes a new aspect of the
model. The ``environment" here may be thought of as a distribution
of a very large, or even infinite number of final states into which
the initial state of the system decays. The Lee-Friedrichs model
\cite{Lee} in Lee's construction was formulated in terms of
non-relativistic quantum field theory using a Hamiltonian for which
the interaction $V^{\rm Lee}\propto
a_Na^{\dag}_Va^{\dag}_\Theta+h.c.$ included an annihilation operator
$a_N$ for the original (unstable) state $N$, multiplied by creation
operators $a^{\dag}_Va^{\dag}_\Theta$ for the two body final state
$V$ and $\Theta$. These operators could have been constructed to
include creation of a many-body environment as well, maintaining its
equivalence with the Friedrichs quantum mechanical form, with
spectral coefficients coupling to a bath. In other words, the decay
process may include not just the specified final states of the decay
model, but also an environment. It is instructive to think of the
spontaneous emission process \cite{CT,Scully} as the most well-known
illustration of a decay into an infinity of final states, where an
excited atomic state decays into a distribution of Fock space states
of the
radiation field (i.e. photons). \\
%It is the purpose of this work to
%show the analogy between the continuum of the decay products and the
%interaction
%of a particle with an environment.\\

 In what follows we show that by associating the continuous spectrum with an environment
the reduced evolution of the discrete states subspace is defined by
a spectral correlation matrix $\alpha(t)$, which is a function
unifying the details of the interaction. The notion of the spectral
correlation matrix is inspired by the well-studied
particle$\otimes$environment theories \cite{Weiss}. In these models
the reduced evolution of the particle is obtained by tracing out the
environmental degrees of freedom, leaving a {\it stochastic}
dynamical equation which invokes the so-called environmental (in
some examples complex) noise $z(t)$. As shown in \cite{Vega}, the
environmental correlation function corresponds to the
autocorrelation value of this noise, i.e. $\alpha(t)=\langle
z^*(t)z(0)\rangle$, and in agreement with the
fluctuation-dissipation theorem turns out to be a time dependent
memory-kernel for the particle energy dissipation.
% We shall stress
%that in the case of our interest, i.e. the Lee-Friedrichs model,
%even though the defined below correlation matrix $\alpha(t)$
%evidently has an analogous structure, an analogous meaning, and
%plays an analogous role in the dynamics of the reduced subspace,
%there is no rigorous evidence for its stochastic interpretation.
%Nevertheless,
Exploiting the analogous properties of the environmental correlation
function and the spectral correlation matrix we investigate the
validity of the non-trivial semigroup evolution for the many channel
decay and identify the necessary conditions with the well-known
Markovian limit \cite{Weiss,CT}. In this way we consider a natural
imbedding of the Wigner-Weisskopf idea into a theory of interaction
with a reservoir (e.g., generalization of Anderson and Fano
\cite{Mahan} and Lee's formulation).
\\

The next two sections include a review of essential results from the
Wigner-Weisskopf theory and the derivation of the Markovian limit,
provided with a brief summary of situations where this limit may be
realized exactly or approximately (readers familiar with these
concepts are welcome to quickly leaf through). In the fourth
section, which includes the main novelty of the paper, we discuss
the effect of the Markovian limit on the Wigner-Weisskopf pole
approximation, and explain how the semigroup evolution law is
achieved for the many channel decay as well.\\

\section{The Wigner-Weisskopf Method}

\begin{figure}[h]
\centering
\includegraphics[width=6.3cm,height=4.5cm]{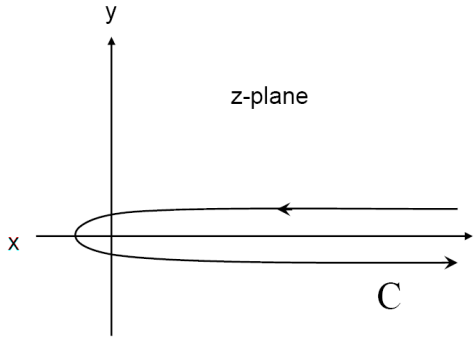}%[width=8cm,height=5cm,angle=180]
\caption{The inverse Laplace transform contour $C$ in
Eq.~(\ref{Ured}).\\} \label{fig1}
\end{figure}

In this section we summarize a functional formulation of the
Wigner-Weisskopf pole approximation theory \cite{WW} and its
generalization to many channel decay \cite{SH}. Following the usual
model for the decay of an unstable system we consider a Hamiltonian
of the form
\begin{equation}
H=H_0+V,\label{H} \end{equation}
where the spectrum of $H_0$ consists of a finite number $N$ of
discrete eigenvalues $\left\{\lambda_\alpha\right\}$, embedded into
a continuum $\left\{\lambda\geq0\right\}$ with spectral weight
$dE(\lambda)=|\lambda\rangle\langle\lambda| d\lambda$. We study, in
particular, the Lee-Friedrichs model \cite{Lee}, for which the
interaction $V$ couples the discrete states to the continuum, but
does not couple continuum states or discrete states among
themselves. The fact that the continuum subspace is associated with
the products of the decay process presents an opportunity for the
introduction of an environment. We assume that the initial unstable
state of the system is given by a superposition of the discrete
eigenstates $\left\{|\phi_{\alpha}\rangle\right\}$ of $H_0$:
\begin{equation}
|\psi_0\rangle=\sum_{\alpha=1}^N c_\alpha|\phi_\alpha\rangle;
~~~H_0|\phi_\alpha\rangle=\lambda_\alpha|\phi_\alpha\rangle.\label{psi0}
\end{equation}
%
%The evolution of the total system, restricted to the subspace of the
%particle spanned by the $N$ discrete eigenvectors
%$\left\{\psi_{\alpha}\right\}$, also called the reduced evolution,
%is governed from the total unitary evolution by the projection on
%the particle subspace:
The reduced evolution, i.e. the evolution of the discrete subspace,
is governed by the reduced propagator $R(z)$. This propagator in
Laplace transform is defined (for ${\rm Im}~z>0$) by
projection %of the Laplace transform $t\rightarrow z$ (for $Im~z>0$)
of the total system propagator
\begin{equation}
U(z)=\int_{0}^\infty dt~{\rm e}^{izt}{\rm
e}^{-iHt}=\frac{i}{z-H}\label{Uz}
\end{equation}
on the discrete subspace. The $\alpha\beta$ matrix element of the
reduced propagator $R(z)$ in the Laplace domain is
\begin{equation}
R_{\alpha\beta}(z)\equiv-i\langle
\phi_\alpha|U(z)|\phi_\beta\rangle=\langle
\phi_\alpha|\frac{1}{z-H}|\phi_\beta\rangle.\label{Rab}
\end{equation}
Eq.~(\ref{Rab}) may be written using $N\times N$ matrix notation
(defining $W(z)$)
\begin{equation}
R(z)=\frac{1}{z-W(z)},\label{R} \end{equation}
which is confined to the discrete subspace, and where, by comparing
with Eq.~(\ref{Uz}), $W(z)$ is understood as the Laplace space
reduced evolution generator. The reduced evolution in the time
domain, dictated by $U^{red}(t)$, is given by the inverse Laplace
transform of Eq.~(\ref{R})
\begin{equation}
U^{red}(t)=\frac{1}{2\pi i}\int_C R(z){\rm e}^{-izt} dz,\label{Ured}
\end{equation}
where the contour of the integration $C$, shown in Fig.~\ref{fig1},
runs slightly above the real line on the $z$-plane from $+\infty$ to
zero (the bottom of the positive spectrum), and then, around the
branch point, from zero back to $+\infty$ slightly below the real
line.

Since, in the general case, the exact calculation of
Eq.~(\ref{Ured}) is difficult, one is interested in a useful
approximation. We first note that there can be {\it no pole} for
${\rm Im}~z\neq0$ in Eq.~(\ref{R}) on the first Riemann sheet (see
Appendix). However, we may explicitly continue the integration in
Eq.~(\ref{Ured}) analytically to the second
Riemann sheet (using Eq.~(\ref{Rab})) \cite{SH}. Doing so yields %This may be achieved
%by constructing the analytical continuation of Eq.~(\ref{R}),
%extending it to the second Riemann sheet ($Im~z<0$) by
%
\begin{equation}
R^{\rm II}(z)=\frac{1}{z-W^{\rm II}(z)},\\\label{R2}
\end{equation}
\begin{figure}[h]
\centering
\includegraphics[width=6cm,height=4.7cm]{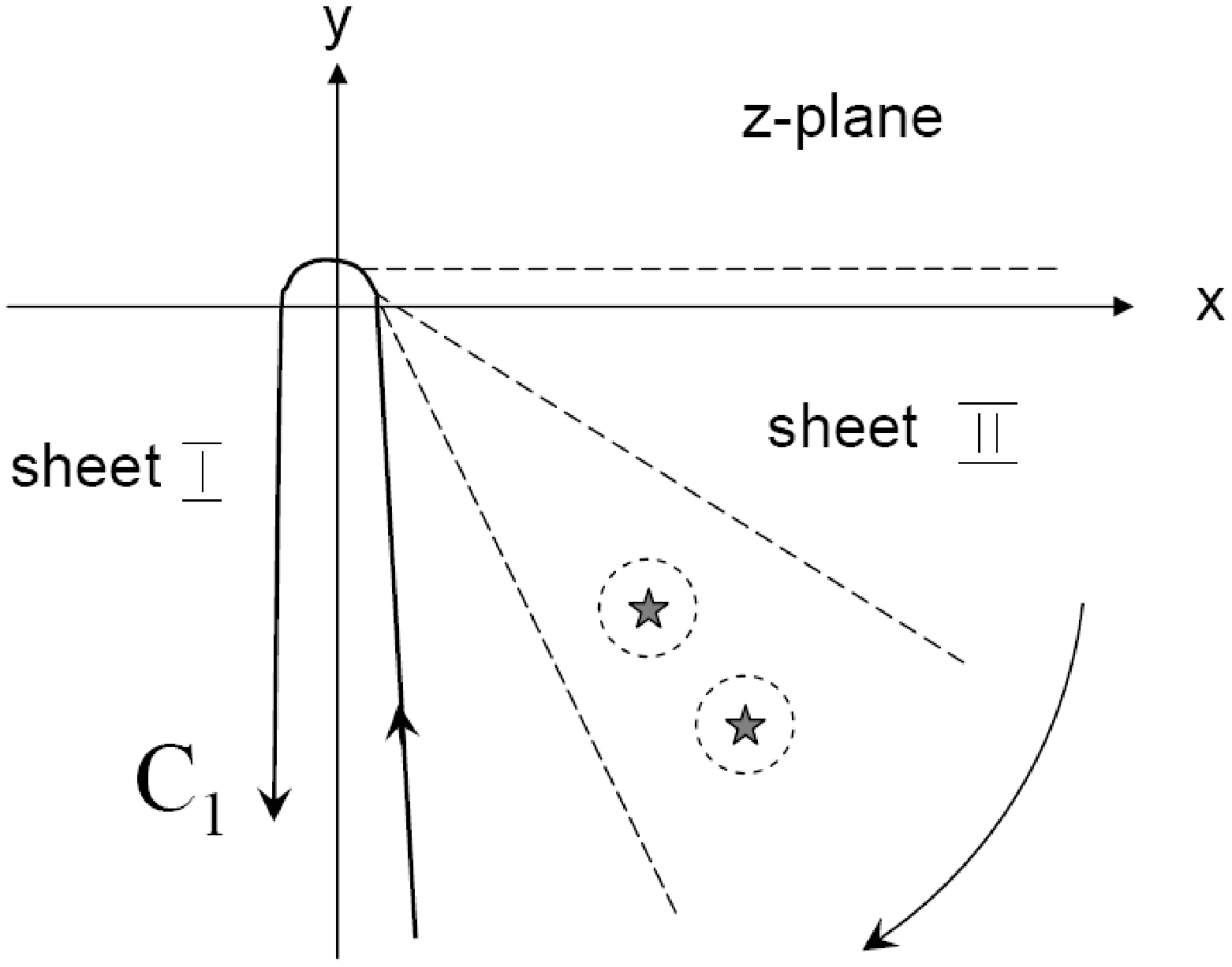}%[width=8cm,height=5cm,angle=180]
\caption{The modified inverse Laplace transform contour $C_1$ in
Eq.~(\ref{UredWW}). The poles are denoted schematically by the
stars.} \label{fig2}
\end{figure}
which allows one to deform the contour of integration. The
eigenvalues of $W^{\rm II}(z)$ (see below) determine the poles of
$R^{\rm II}(z)$ in the lower half plane. The procedure, described in
detail in \cite{SH}, results in the following alternative expression
for the reduced propagator
\begin{equation}
\!\!U^{red}(t)\!\!=\!\! \frac{1}{2\pi i}\!\!\int_{C_1} R^{\rm I,
II}(z){\rm e}^{-izt} dz-2\pi i\sum_j{\rm e}^{-iz_j t}Res\left[R^{\rm
II}(z_j)\right],\label{UredWW}
\end{equation}
where $Res[R^{II}(z_j))]$ ($R^{II}(z)$ on the right and $R(z)$,
which we denote above by $R^{I}(z)$, on the left) is the residue of
$R^{II}(z)$ at the pole position $z_j$ and the contour of the
integration $C_1$, shown in Fig.~\ref{fig2}, now runs around the
branch point along the negative
imaginary axis. %By such a restatement of the original integral
%appearing in the right hand-side of Eq.~(\ref{Ured}) we yield the
%following observation:
The integration along the contour $C_1$ carries the factor ${\rm
e}^{-izt}$ for $z$ in the lower half plane. Hence, for $t>0$ and not
too small$^{1}$\footnotetext[1]{The scale of the value necessary to
go beyond the non-exponential region of decay curve, often called
the Zeno time \cite{Misra}, is determined by the dispersion of the
Hamiltonian, as discussed in \cite{Bleistein} and below
(Eqs.~(37),(38)). The onset of the exponential behavior generally
occurs after the curve of steepest descent, which rotates clockwise
in time \cite{Bleistein}, passes the first pole, which then
dominates the time dependence.} one can consider neglecting this
term, called the ``background contribution". (There is a very long
time contribution from the neighborhood of the branch point which we
do not consider here). Doing so, the evaluation of the reduced
propagator reduces to the summation of the contributions of the
residues of the poles of $R^{\rm II}(z)$ in the lower half plane.
The assumed dominance of the these contributions
is called the ``pole approximation".\\

To obtain an expression describing the residues of $R^{\rm II}(z)$,
we focus on the reduced generator $W^{\rm II}(z)$. Even though
$W^{\rm II}(z)$ is not generally self adjoint, Eq.~(\ref{R2}) may be
represented in a way analogous to the spectral theorem, as the sum
of normalized projectors $\left\{Q_\alpha(z)\right\}$, so that
\begin{equation}
R^{\rm
II}(z)=\sum_{\alpha=1}^N\frac{Q_\alpha(z)}{z-\omega_\alpha(z)},\label{RvsQ}
\end{equation}
where
\begin{equation}
Q_\alpha(z)=\frac{|\alpha,z\rangle_{R~L}\langle\alpha,z|}{_L\langle\alpha,z|\alpha,z\rangle_R}
\label{Qa}
\end{equation}
are made of the left and right eigenvectors of $W^{\rm II}(z)$
(corresponding to appropriate linear combinations of the
eigenvectors of $H_0$, sometimes called "decay eigenstates"):

$$
W^{\rm
II}(z)|\alpha,z\rangle_R=\omega_\alpha(z)|\alpha,z\rangle_R,$$
\begin{equation}
_L\langle\alpha,z|_RW^{\rm
II}(z)=\omega_\alpha(z)_L\langle\alpha,z|. \label{eignW}
\end{equation}
Clearly,
$$
_L\langle\beta,z|W^{\rm
II}(z)|\alpha,z\rangle_R=\omega_\alpha(z)_L\langle\beta,z|\alpha,z\rangle_R
=$$\begin{equation}=\omega_\beta(z)_L\langle\beta,z|\alpha,z\rangle_R
\end{equation}
may be valid for $\omega_\alpha(z)\neq\omega_\beta(z)$ (for any $z$)
only if
\begin{equation}
_L\langle\beta,z|\alpha,z\rangle_R=0.\label{or}
\end{equation}
This orthogonality relation for the eigenvectors of $W^{\rm II}(z)$
provides the orthogonality of the appropriately normalized
projectors
\begin{equation}
Q_\alpha(z)Q_\beta(z)=Q_\alpha(z)\delta_{\alpha\beta},\label{OR}
\end{equation}
{\it at each  point $z$.}  Eq.~(\ref{RvsQ}) follows from
Eq.~(\ref{R2}) by using the spectral representation $W^{II}(z) =
\sum_\alpha^N
\omega_\alpha(z)|\alpha,z\rangle_{R~L}\langle\alpha,z|$ and
the orthogonality properties of the $Q_\alpha(z)$.\\% to sum the
%series expansion of the expression (\ref{R2}),  i.e., the same
%theorem holds for functions of such finite non-Hermitian matrices as
%for the spectral representation of self-adjoint operators in
%functional analysis on a general Hilbert
%space.\\

Eq.~(\ref{OR}) plays crucial role in examination of the semigroup
property of the reduced evolution. Applying the pole approximation
procedure, we neglect the ``background contribution" and approximate
the reduced propagator by the sum of the residues of $R^{\rm II}(z)$
Eq.~(\ref{RvsQ}), which for weak coupling $V$ may be well
approximated \cite{SH} to yield
\begin{equation}
U^{red}(t)\cong \sum_j{\rm e}^{-iz_j
t}Q_{\alpha_j}(z_j),\label{Ured1}
\end{equation}
where $\alpha_j$ corresponds to the singularity at
$z_j=\omega_{\alpha_j}(z_j)$. Repeated application of this reduced
evolution is then
\begin{equation}
U^{red}(t_2)U^{red}(t_1)\cong\sum_{j,k}{\rm e}^{-iz_j t_2}{\rm
e}^{-iz_k t_1}Q_{\alpha_j}(z_j)Q_{\alpha_k}(z_k).\label{U2U1}
\end{equation}
Although for the single channel problem, if there is just one pole,
the projectors product in the right hand-side of the last equation
is trivially unity and Eq.~(\ref{U2U1}) shows semigroup decay, for
the many channel decay with many poles the projectors
$Q_{\alpha_j}(z_j), Q_{\alpha_k}(z_k)$ are generally evaluated at
{\it different} pole locations on the Laplace plane. For $z_j\neq
z_k$ the orthogonality relation Eq.~(\ref{OR}) can no longer ensure
$Q_{\alpha_j}(z_j)Q_{\alpha_k}(z_k)=0$. {\it Thus, even in the pole
approximation, the semigroup evolution is generally not valid}
\cite{CH,MH}, i.e.
\begin{equation}
U^{red}(t_2)U^{red}(t_1)\neq U^{red}(t_1+t_2)=\sum_j {\rm e}^{-iz_j
(t_1+t_2)}Q_{\alpha_j}(z_j).\label{U1+2}
\end{equation}
%
%If there were only a single pole $z_\alpha$, as may occur in the
%case of the single channel decay, semigroup evolution would be a
%very good approximation. But already for two different poles
%$z_\alpha\neq z_\beta$ Eq.~(\ref{U2U1}) is not generally a
%semigroup.
In spite of this conclusion, many experiments display semigroup
decay to high accuracy \cite{Fermilab}, while the estimates
\cite{CH,MH} have shown that the deviations predicted by the
Wigner-Weisskopf theory
would exceed the experimental error \cite{Fermilab}.\\

%\begin{widetext}
\section{Association of the spectral density function with an environment}

In this section we sketch a treatment for the reduced system
dynamics familiar in the field of quantum optics \cite{CT,Scully}
and condensed matter physics \cite{Weiss,Mahan}.
%and quantum electrodynamics \cite{CT}.
We adopt the Lee-Friedrichs model of the system described in the
previous section for the total Hamiltonian given by Eq.~(\ref{H}).
For our current purpose it is not necessary to make any preliminary
assumptions (other than, for simplicity, degeneracy) regarding the structure of the continuous part of $H_0$; it may be bounded from below or may not. % We remind that the
%following derivations are restricted to the Lee-Friedrichs model,
%which assumes that the interaction $V$ couples only between the
%discrete states and the continuum, but does not couple any of
%discrete or any of continuous states among themselves.
The interaction picture propagator, defined through the total system
propagator $U(t_2-t_1)$ and the unperturbed total system propagator
$U_0(t-t_0)={\rm e}^{-iH_0(t-t_0)}$ is given by
\begin{equation}
\tilde{U}(t_2,t_1)=U_0^{-1}(t_2-t_0)U(t_2-t_1)U_0(t_1-t_0)\label{tildeU}
\end{equation}
and obeys the equation
\begin{equation}
i\frac{d}{dt}\tilde{U}(t,t_0)=\tilde{V}(t)\tilde{U}(t,t_0),\label{dttildeU}
\end{equation}
where
\begin{equation}
\tilde{V}(t)=U_0^{-1}(t-t_0)VU_0(t-t_0)\label{Vt}
\end{equation}
is the interaction picture Hamiltonian. Integrating
Eq.~(\ref{dttildeU}) and iterating it one time we get the exact
equation
\begin{equation}
\!\!\tilde{U}(t,t_0)=1-i\!\!\int_{t_0}^t\tilde{V}(\tau)
d\tau-\!\!\int_{t_0}^t\int_{t_0}^\tau\tilde{V}(\tau)
\tilde{V}(\tau')\tilde{U}(\tau',t_0)d\tau' d\tau.\label{Vint}
\end{equation}
Projecting the last expression on the discrete subspace we find
$$
\tilde{U}^{red}_{\alpha\beta}(t,t_0)\equiv\langle
\phi_\alpha|\tilde{U}(t,t_0)|\phi_\beta\rangle=$$%
%\delta_{\alpha\beta}+i\int_{t_0}^t\langle
%\phi_\alpha|\tilde{V}(\tau)|\phi_\beta\rangle d\tau-$$$$\!\!-
%\sum_\gamma\int_{t_0}^t\!\!\int_{t_0}^\tau\!\!\int_\lambda\langle
%\phi_\alpha|\tilde{V}(\tau)|\lambda\rangle\langle \lambda|
%\tilde{V}(\tau')|\phi_\gamma\rangle\tilde{U}^{red}_{\gamma\beta}(\tau'-t_0)\
%d\lambda d\tau' d\tau=$$
$$
=\delta_{\alpha\beta}-
\sum_\gamma\int_{t_0}^t\int_{t_0}^\tau\int_\lambda{\rm
e}^{i\lambda_\alpha(\tau-t_0)}\langle \phi_\alpha|V
|\lambda\rangle{\rm e}^{-i\lambda(\tau-t_0)}\times$$
\begin{equation}\times{\rm
e}^{i\lambda(\tau'-t_0)}\langle \lambda| V|\phi_\gamma\rangle{\rm
e}^{-i\lambda_\gamma(\tau'-t_0)}\tilde{U}^{red}_{\gamma\beta}(\tau',t_0)
d\lambda d\tau' d\tau.\label{PVint}
\end{equation}
Here the first order term $\langle
\phi_\alpha|\tilde{V}(\tau)|\phi_\beta\rangle$ has vanished, because
by assumption, $V$ does not couple the discrete states among
themselves, and the last term was obtained using Eq.~(\ref{Vt}).
Next we differentiate Eq.~(\ref{PVint}) with respect to $t$ and
obtain the {\it reduced}, i.e., projected into the unstable
subspace, master equation

$$
\frac{d}{dt}\tilde{U}^{red}_{\alpha\beta}(t,t_0)=- \sum_\gamma{\rm
e}^{i\lambda_\alpha (t-t_0)}\times$$
\begin{equation}\times\int_{t_0}^t \int_\lambda{\rm
e}^{-i\lambda (t-\tau')}\omega_{\alpha\gamma}(\lambda)~d\lambda{\rm
e}^{-i\lambda_\gamma(\tau'-t_0)}\tilde{U}^{red}_{\gamma\beta}(\tau',t_0)
d\tau' ,\label{DtR}
\end{equation}
where the matrix elements
\begin{equation}
\omega_{\alpha\gamma}(\lambda)\equiv\langle\phi_\alpha|V|\lambda\rangle\langle\lambda|V|\phi_\gamma\rangle\label{SDFb}
\end{equation}
form the so-called {\it spectral density matrix}. The integral
transform of the latter
\begin{equation}
\tilde{\alpha}_{\alpha\gamma}(t,\tau')\equiv{\rm e}^{i\lambda_\alpha
(t-t_0)}\left[\int {\rm e}^{-i\lambda
(t-\tau')}\omega_{\alpha\gamma}d\lambda\right] {\rm
e}^{-i\lambda_\gamma(\tau'-t_0)}\label{tildalpha}
\end{equation}
defines the elements of the {\it spectral correlation matrix} within
the interaction representation. Using the last expression,
Eq.~(\ref{DtR}) may be written as
\begin{equation}
\frac{d}{dt}\tilde{U}^{red}_{\alpha\beta}(t,t_0)=-\sum_\gamma\int_{t_0}^t
\tilde{\alpha}_{\alpha\gamma}(t,\tau')\tilde{U}^{red}_{\gamma\beta}(\tau',t_0)
d\tau' .\label{DtR1}
\end{equation}
Transforming back to the Schr\"{o}dinger representation, we use the
inverse version of Eq.~(\ref{tildeU}). Taking into account that the
spectral correlation matrix $\tilde{\alpha}(t,\tau')$ transforms
analogously to $\tilde{U}(t_2,t_1)$, that
$\frac{d}{dt}\tilde{U}(t,t_0)=iH_{0}U_0^{-1}(t-t_0)U(t,t_0)
+U_0^{-1}(t-t_0)\frac{d}{dt}U(t,t_0)$, and that $H_0$ and
$U_0(t_2-t_1)$ are diagonal in the reduced basis
$\left\{|\phi_\alpha\rangle\right\}$ representation, in terms of
matrix notation we obtain
\begin{equation}
\!\!\frac{d}{dt}U^{red}(t,t_0)\!=\!-
iH_{0}^{red}U^{red}(t,t_0)-\!\!\int_{t_0}^t\!\!\alpha(t-\tau')U^{red}(\tau',t_0)
d\tau'.\label{DotR}
\end{equation}
Here
\begin{equation}
H_{0}^{red}=\sum_{\alpha=1}^N\lambda_\alpha|\phi_\alpha\rangle\langle\phi_\alpha|\label{H0S}
\end{equation}
is the discrete part of the unperturbed Hamiltonian $H_0$, and the
spectral correlation matrix defined as
\begin{equation}
\alpha(t)\equiv\int_\lambda{\rm e}^{-i\lambda
t}\omega(\lambda)~d\lambda\label{alphaI}
\end{equation}
is the Fourier transform of the spectral density matrix
%the integral transform of the spectral density matrix $\omega(\lambda)$ (which elements are given by Eq.~(\ref{SDFb})) %The integral transform
%of $\omega(\lambda)$, involved in the right hand-side of
%Eq.~(\ref{DtR}) (here and in Eq.~(\ref{DtR}) above the overall
%exponential factor ${\rm e}^{i(\lambda_\alpha-\lambda_\gamma)
%(t-t_0)}$ is, as tilde denotes, due to the interaction
%representation)
%
\begin{equation}
\omega(\lambda)\equiv\sum_{\alpha,\gamma=1}^N
\omega_{\alpha\gamma}(\lambda)|\phi_\alpha\rangle\langle\phi_\gamma|.\label{hatomega}
\end{equation}
We argue that $\alpha(t)$ defined by Eq.~(\ref{alphaI}) can be
associated with the ``noisy" environmental correlation function
$\alpha(t)=\langle z^*(t)z(0)\rangle$ mentioned earlier, since the
microscopic definition \cite{Weiss,Vega} of the latter evidently
coincides with Eq.~(\ref{alphaI}) up to an obvious generalization.
The correlation matrix $\alpha(t)$ Eq.~(\ref{alphaI}) represents all
the microscopic details of the interaction, whose properties
determine the type of the reduced evolution, as will be clear from the following. \\

\subsection{Equivalence of Exact Markovian Coupling and Globally Flat and Unbounded Spectral Density Matrix}

Note, that in an ideal case for which
$\omega_{\alpha\gamma}(\lambda)$ Eq.~(\ref{SDFb}) are independent of
$\lambda$ and the continuous spectrum of $H_0$ is unbounded (a
physical example of such a situation could occur for a Stark type
interaction with a bath, which induces a shot noise), the
correlation matrix Eq.~(\ref{alphaI}) reduces to a delta function of
time:
\begin{equation}
\alpha(t-\tau)=\omega \!\!\int_{-\infty}^\infty
{\rm e}^{-i\lambda(t-\tau)}d\lambda%=2\pi\omega_{\alpha\gamma}{\rm e}^{i\lambda_\alpha (t-t_0)}{\rm e}^{-i\lambda_\gamma(\tau'-t_0)}\delta(t-\tau')
=2\pi\omega\delta(t-\tau).\label{MECF}
\end{equation}
Substituting this into Eq.~(\ref{DotR}) we find
\begin{equation}
\frac{d}{dt}U^{red}(t-t_0)=\left(-iH_{0}^{red}-
2\pi\omega\right)U^{red}(t-t_0).\label{Mdtr}
\end{equation}
The last equation is a {\it local in time first order differential
equation with constant evolution generator}, i.e. an equation
describing semigroup evolution, also called the Markov equation,
since it describes a Markovian stochastic process \cite{vanKampen}.
Its solution for $t>t_0$ is
\begin{equation}
U^{red}(t-t_0)={\rm e}^{(-iH_{0}^{red}-\Gamma)(t-t_0)},\label{MRb}
\end{equation}
where
the {\it decay matrix} $\Gamma$ is given by
%
%\begin{equation}
%\Gamma=\int_{t_0}^t \alpha(\tau) d\tau=\int_{t_0}^t 2\pi\omega\delta(t-\tau')d\tau=2\pi\omega.\\\label{Gamma}
%\end{equation}
%%
%
\begin{equation}
\Gamma \equiv\int_{t_0}^t \alpha(\tau) d\tau=\int_{t_0}^t
2\pi\omega\delta(t-\tau')d\tau= 2\pi\omega.\\\label{Gammab}
\end{equation}
The well-known demonstration of the Zeno effect for very short times
$t$ relies on an expansion of the survival probability $P(t)$ of a
unstable state $|\psi\rangle$ in a series \cite{Bleistein}
$$
P(t)=|\langle\psi|{\rm
e}^{-iHt}|\psi\rangle|^2\simeq$$\begin{equation}
\simeq|\langle\psi|1-iHt-\frac{1}{2}H^2t^2+\dots|\psi\rangle|^2\simeq1-t^2\Delta
H^2\label{Ps}
\end{equation}
where
\begin{equation}
\Delta H^2=\langle\psi|H^2|\psi\rangle
-\left(\langle\psi|H|\psi\rangle\right)^2\label{DH} \end{equation}
is the dispersion of the total Hamiltonian $H$ in the state
$|\psi\rangle$. The standard argument leading one to a conclusion
about the principle impossibility of pure semigroup evolution is the
presumable possibility to cut off the above expansion after the
second order in $t$. Note, however, that for the Lee-Friedrichs
model in the Markovian limit the coefficient $\Delta H^2$ diverges.
If, for example, $|\psi\rangle=|\psi_\alpha\rangle$, Eq.~(\ref{DH})
becomes
\begin{equation}
\left(\Delta
H^2\right)^{Markov}\!\!\!=\langle\phi_\alpha|V^2|\phi_\alpha\rangle
= \int_{-\infty}^{\infty} d\lambda\omega_{\alpha\alpha}(\lambda)
\Rightarrow \infty\label{MDH}
\end{equation}
Hence, the truncation of the expansion is invalid together with the
physically interpreted conclusion. It should be stressed, that under
our assumptions regarding the structure of the coupling, there is no
Zeno (see also \cite{Adler}) effect and the result Eq.~(\ref{MRb})
is exact.\\ %Note, that this argument would be valid for the
%semibounded
%or slowly varying case as well.\\

Now we briefly review how such assumptions may be realized. For this
purpose we adopt the usual approach to the original definition of
the spectral density function, i.e. as the continuum limit of a
quasi-continuous spectrum. We assume, that the unperturbed system,
described by $H_0$, is confined to a large box with some standard
boundary conditions. The spectrum of $H_0$ then consists of $N$
discrete eigenstates embedded into a quasi-continuous spectrum of
the environment including the decay products. Each element
$\omega_{\alpha\gamma}(\lambda)$ of the spectral density matrix
$\omega(\lambda)$ Eq.~(\ref{hatomega}) is defined as the product of
the interaction amplitudes
$\langle\phi_\alpha|V|\lambda\rangle\langle\lambda|V|\phi_\gamma\rangle$,
weighted by the local degeneracy of the environmental states
\begin{equation}\omega_{\alpha\gamma}(\lambda)=
\langle\phi_\alpha|V|\lambda\rangle\langle\lambda|V|\phi_\gamma\rangle
D(\lambda),\label{gD}
\end{equation}
where the {\it density of states} $D(\lambda)$ serves as an
effective ``coarse graining". When the spectrum is truly continuous
$D(\lambda)$ may be absorbed in
$\langle\phi_\alpha|V|\lambda\rangle\langle\lambda|V|\phi_\gamma\rangle
d\lambda$. The interaction amplitudes
$\langle\phi_\alpha|V|\lambda\rangle\langle\lambda|V|\phi_\gamma\rangle$
are determined by the microscopic details, so their functional
dependence on $\lambda$ is dictated by the nature of the
interaction. On the other hand, the density of the environmental
states $D(\lambda)$ is determined by $H_0$ and, in particular, by
the {\it boundary conditions} to which the system is confined. Thus,
$D(\lambda)$ is at our disposal. Designing the geometry of the space
we may find a $D(\lambda)$ which suppresses, at least approximately,
the $\lambda$-dependence of
$\langle\phi_\alpha|V|\lambda\rangle\langle\lambda|V|\phi_\gamma\rangle$,
and yields a semigroup evolution with a desirable
accuracy$^{2}$\footnotetext[2]{There exists a class of
phenomenological descriptions of the spectral density functions
$\omega_{\alpha\gamma}(\lambda)\propto\eta_s\lambda_c^{1-s}\lambda^s{\rm
e}^{-\lambda/\lambda_c}$ \cite{Weiss}, where $\eta_s$ is the
viscosity constant, the exponential factor provides a smooth cut-off
modulated at the frequency $\lambda_c$, and where $0<s<1$ and $s>1$
describe the so-called sub-ohmic and the super-ohmic interaction
respectively. The slower is the dependence of the spectral density
function on $\lambda$, the closer is the reduced evolution to the
Markovian limit. The boundary case of $s=1$ corresponds to the ohmic
interaction, as for example, the dipole interaction of a particle
with the {\it free} electromagnetic radiation field, which is still
associated with an {\it approximate}
Markovian spectral coupling.}.\\%

\subsection{Approximate Markovian Coupling vs. Weak Interaction and Resonance.}

In the preceding subsection we have assumed a continuous spectrum on
the whole real line in order to obtain an exact semigroup evolution.
However, if there is a resonance and the coupling is weak, the
leading behavior of the system is dictated by small regions of the
continuous spectrum where the effect of the interaction is sharply
enhanced. Such regions occur at $\lambda$ sufficiently close to the
resonances, which for small coupling are close to the eigenvalues
$\left\{\lambda_\alpha\right\}$ of the discrete subspace. To show
this we return to Eq.~(\ref{DtR})
%
%$$
%\frac{d}{dt}\tilde{U}^{red}_{\alpha\beta}(t,t_0)=-
%\sum_\gamma\int_{t_0}^t{\rm e}^{i\lambda_\alpha (t-t_0)} {\rm
%e}^{-i\lambda_\gamma(\tau'-t_0)}\times$$$$\times\int_\lambda{\rm
%e}^{-i\lambda
%(t-\tau')}\omega_{\alpha\gamma}(\lambda)~d\lambda\tilde{U}^{red}_{\gamma\beta}(\tau'-t_0)
%d\tau' ,~~~~~\mbox{$(Eq.~23)$}$$
%%
and see whether it is possible to utilize it approximately, even if
$\omega(\lambda)$ is not constant and the continuous part of the
$H_0$ spectrum is bounded from below. Inspecting the integrand of
the last term on the right hand-side of Eq.~(\ref{DtR}) we note that
the collapse of the memory kernel may result from rather symmetrical
manipulations with respect to the spectral or the time variables.
For the perfectly Markovian coupling $\omega(\lambda)$ is constant
and unbounded, the integral over $\lambda$ yields $\delta(t-\tau')$
and admits the semigroup evolution of Eq.~(\ref{Mdtr}). Conversely,
if it is justified to neglect the difference between
$\tilde{U}^{red}_{\gamma\beta}(\tau',t_0)$ and
$\tilde{U}^{red}_{\gamma\beta}(t,t_0)$ and to stretch the limits of
the integration over $\tau'$ to $\pm\infty$, it also yields a
Markovian equation of the form of Eq.~(\ref{MECF}) (and results in a
factor $\delta(\lambda-\lambda_\gamma)$ as well). The error caused
by the replacement
$\tilde{U}^{red}_{\gamma\beta}(\tau',t_0)\rightarrow
\tilde{U}^{red}_{\gamma\beta}(t,t_0)\propto{\cal O}(V^2)$ is
negligible in case the interaction is sufficiently weak, because the
last term of Eq.~(\ref{DtR}) is already second-order in $V$. Such an
approximation yields a solution, which up to the second order in $V$
is exact, and provides a direct interpretation of the original
perturbative computation of Weisskopf and Wigner \cite{WW} in the
resolvent formalism. Making the weak coupling assumption, we {\it
approximate} the reduced master equation (\ref{DtR}) by
$$
\frac{d}{dt}\tilde{U}^{red}_{\alpha\beta}(t,t_0)=-
\sum_\gamma\int_\lambda{\rm e}^{i\lambda_\alpha (t-t_0)}{\rm
e}^{i\lambda_\gamma t_0}{\rm e}^{-i\lambda t}
\omega_{\alpha\gamma}(\lambda)\times$$
\begin{equation}\times\int_{t_0}^{t}{\rm
e}^{-i(\lambda_\gamma-\lambda)\tau'}
 d\tau'd\lambda~\tilde{U}^{red}_{\gamma\beta}(t,t_0),\label{DtRtau}\end{equation}
and focus on the integration over the time. Since the interaction
$V$ is time independent, one may select the time origin such that
$t_0=-T/2$ and $t=T/2$. Then the integration
 over $\tau'$ in Eq.~(\ref{DtRtau}) yields \cite{CT} %(p. 26)
\begin{equation}
\int_{-T/2}^{T/2}{\rm e}^{-i(\lambda_\gamma-\lambda)\tau'}d\tau'=\frac{1}{\pi}
\frac{\sin\left[(\lambda-\lambda_\gamma)T/2\right]}{(\lambda-\lambda_\gamma)}.\label{Sinc}
\end{equation}
This is the well-known sinc-function, which in the limit
$T\rightarrow\infty$ is one of the definitions of the Dirac
delta-function. The fact that the sinc-function Eq.~(\ref{Sinc}) is
sharply peaked around $\lambda=\lambda_\gamma$, and falls quickly
when $\lambda-\lambda_\gamma>\frac{2\pi}{T}$ gives rise to the
notion of the {\it resonance}: substituted back into the integral
over $\lambda$ in Eq.~(\ref{DtRtau}); the sinc-function suppresses
all the values of the spectral density function
$\omega_{\alpha\gamma}(\lambda)$ outside this region. Thus, we
observe that stretching the limits of integration over $\tau'$ in
Eq.~(\ref{DtRtau}) to infinity implies the replacement of the actual
sinc-function Eq.~(\ref{Sinc}) by the Dirac delta-function
$\delta(\lambda-\lambda_\gamma)$. The error induced by this
approximation is legitimate to neglect for $t=T/2$ sufficiently
large, and then Eq.~(\ref{DtRtau}) becomes the further approximated
reduced master equation
% at the most brutal approximation we
%are allowed to replace $\omega(\lambda)$ by $\omega(\lambda_0)$ and also to stretch the integration over $\lambda$
%to $\pm\infty$
%
\begin{equation}
\!\!\!\frac{d}{dt}\tilde{U}^{red}_{\alpha\beta}(t-t_0)\approx\!%-
%\sum_\gamma\int_\lambda{\rm e}^{i\lambda_\alpha (t-t_0)}{\rm
%e}^{i\lambda_\gamma t_0}{\rm e}^{-i\lambda t}
%\omega_{\alpha\gamma}(\lambda)\times$$$$\times\delta(\lambda-\lambda_\gamma)
%d\lambda~\tilde{U}^{red}_{\gamma\beta}(t-t_0)=$$
% \begin{equation}
- 2\pi\sum_\gamma{\rm e}^{i(\lambda_\alpha-\lambda_\gamma) (t-t_0)}
\omega_{\alpha\gamma}(\lambda_\gamma)\tilde{U}^{red}_{\gamma\beta}(t-t_0).\label{RDtR}\end{equation}
Transforming the last expression back to the Schr\"{o}dinger
representation we regain in matrix notation
Eqs.~(\ref{Mdtr},\ref{MRb}), except that instead of the constant
density matrix $\omega$ we have the ``resonant spectral density
matrix"
\begin{equation}
\omega^{Res}=
\sum_{\alpha,\gamma=1}^N\omega_{\alpha\gamma}(\lambda=\lambda_\gamma)|\phi_\alpha\rangle\langle\phi_\gamma|,
\label{omegaR}
\end{equation}
whose elements are given by $\omega_{\alpha\gamma}(\lambda)$
Eq.~(\ref{SDFb}) evaluated at the discrete eigenvalues
$\lambda=\left\{\lambda_\gamma\right\}$ of $H_0$, where the
subscript $\gamma\in[1,N]$ corresponds to the column index. Thus,
for large enough $t=T/2$ the contribution of the spectral density
over the sharp resonances is very much enhanced, the structure of
the rest of the continuous spectrum of $H_0$ is quite unimportant
for the reduced dynamics, and the error involved in the approximate
Markovian
Eq.~(\ref{RDtR}) is small.\\

From the above considerations it is clear that the
resonance-Markovian assumption fails if: (i) the interaction is
strong, and (ii) $\omega(\lambda)$ vanishes or undergoes significant
changes near the resonant values
$\lambda\cong\left\{\lambda_\gamma\right\}$. %, or if there is no resonant frequency,
%i.e. if the continuous part of $H_0$ has a gap}.
%This, however, is a rather rare event. %On the contrary, if the
%interaction is not very strong and the spectral density matrix
%$\omega(\lambda)$ is smooth enough within very close neighborhood of
%the resonances (i.e. locally ``flat"), then
Nevertheless, the Markovian approximation is usually very good and
suits a large variety of natural environments. To see why, let us
restore the Plank constant $\hbar$, taken so far to
be equal to unity. Doing so shows \cite{CT}% (p. 27)
, that the region of the non-negligible values of the sinc-function,
i.e. the resonance area, is equal to $\frac{2\pi\hbar}{T}$. Since,
for some reasonable $T$, the delta function approximation of the
sinc-function is clearly a good one, the smoothness requirement of
$\omega(\lambda)$ refers to a finite number of small regions in the
spectrum. The requirements of the approximation in Eq.~(\ref{RDtR})
are then not highly restrictive. Therefore, it would be
experimentally difficult to detect
%
%Since $\hbar$ is very small, not only the delta-function
%approximation of the sinc-function is good, but also the smoothness
%requirement of $\omega(\lambda)$ refers to the
%extremely tiny region of the spectrum. %Practically, it turns out
%that
%Due to the factor $\hbar$, the requirements of the approximation in
%Eq.~(\ref{RDtR}) are so non-restrictive that the experimentalists
%have usually to work very hard to detect
any deviation from the semigroup evolution, which occurs only on
``atomic" time-scales. In order to do this, one may need to
deliberately destroy the smoothness of the environmental spectrum
(by careful choice of the boundary conditions), which on such small
scales might be difficult \cite{QED,PGB}.  If $\omega(\lambda)$ is
changing rapidly (or vanishes) near the resonances, the Markovian
approximation would be invalidated and the lifetime of the Zeno
effect would be lengthened.% \cite{Raizen}.

%(Remark: making $\omega(\lambda)$ fast changing
%near the resonances we ruin the Markovian approximation and enlarge the life-time of the Zeno effect).\\

\section{Wigner-Weisskopf Pole Approximation in Markovian Limit}

In the preceding section we saw that the general
integro-differential equation (\ref{DotR}) for the reduced
propagator $U^{red}(t,t_0)$
%
%\begin{equation}
%\frac{d}{dt}R(t-t_0)=-iH_{0}^{red}R(t-t_0)-\int_{t_0}^t\alpha(t-\tau')R(\tau'-t_0)d\tau',\label{mkR}
%\end{equation}
%
%where $H_{0}^{red}$ Eq.~(\ref{H0S}) is the discrete part of the
%unperturbed Hamiltonian, and where the correlation matrix
%$\alpha(t-\tau')$ is the time dependent memory kernel. The last
%equation
expresses the fact that the dynamics of the open system may depend
on its history. Yet, semigroup evolution for the open system can be
achieved, if due to some special circumstances $U^{red}(t,t_0)$ may
reduce to a dynamics approximated by the form of Eq.~(\ref{Mdtr}).
Since, in contrast to the closed system, the energy of the open
system need not be conserved, the reduced evolution generator is not
necessarily Hermitian and may have effective complex eigenvalues
responsible for the decay.\\

As mentioned, Eq.~(\ref{DotR}) may reduce to a Markovian form either
exactly or approximately. The first case occurs when the spectral
density matrix $\omega(\lambda)$ Eq.~(\ref{SDFb}) is constant and
unbounded. The second, much more realistic, yields the desired
effect approximately through a combination of the second order weak
coupling perturbation and resonances. In either case, the dramatic
mutation of the time dependent spectral correlation matrix
$\alpha(t)$ Eq.~(\ref{alphaI}) into a delta-correlated operator,
leads to the equation of the form of Eq.~(\ref{Mdtr}), which {\it
independently of the dimension of the reduced subspace} results in a
semigroup solution of the form of Eq.~(\ref{MRb}), and thus, permits
a semigroup decay also for the many channel problem, as shown
in our analysis of the Wigner-Weisskopf method.\\

To clarify the impact of the Markovian assumption on the pole
approximation approach we first review its basics for the idealized
Markovian limit. Note that in case the continuum spectrum is
unbounded, there is no branch point, and the integration path $C$ in
the fundamental equation (\ref{Ured}) goes above the whole real
line.
\begin{figure}[h]
\centering
\includegraphics[width=6cm,height=4.3cm]{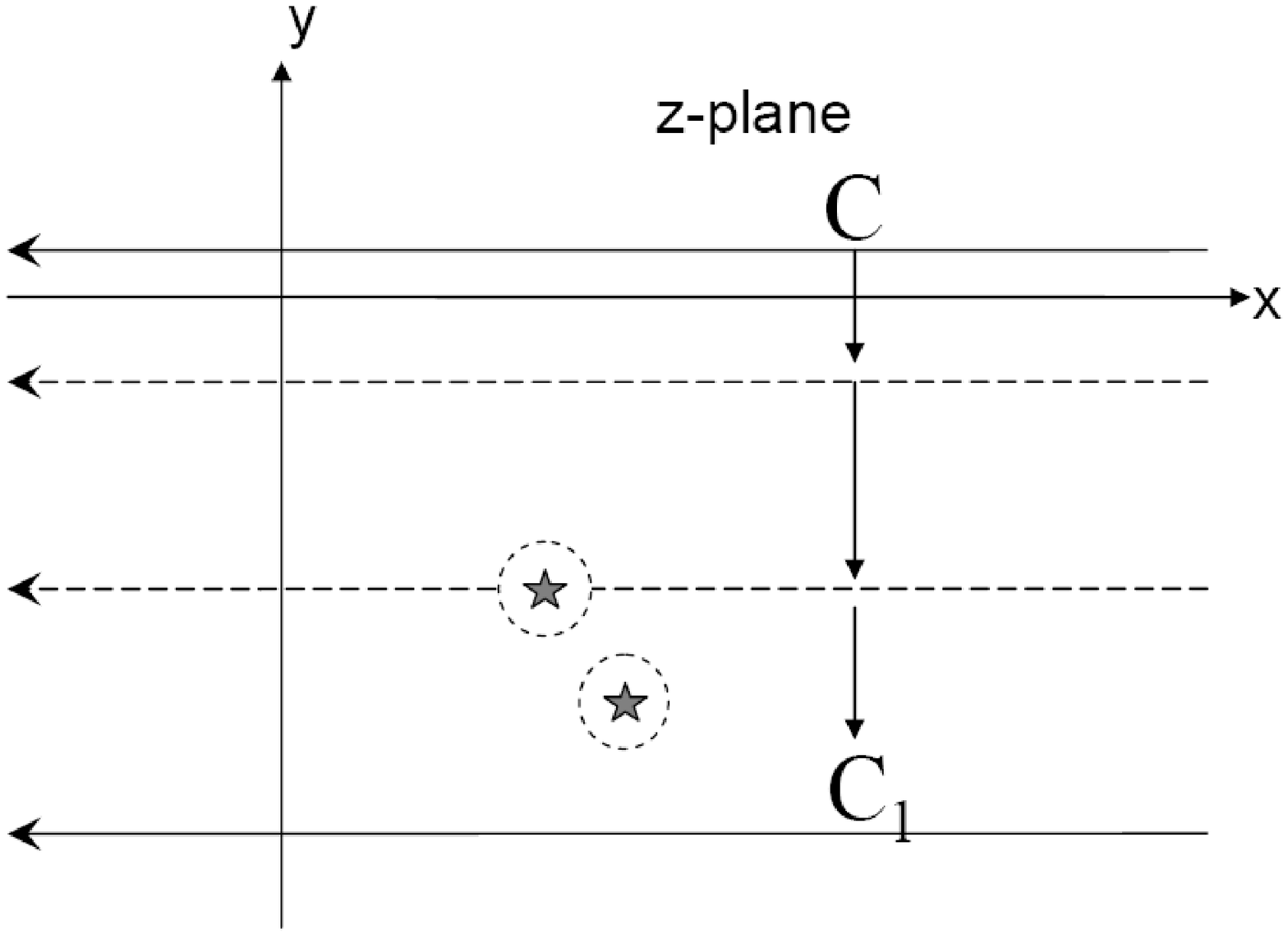}%[width=8cm,height=5cm,angle=180]
\caption{The inverse Laplace transform contour $C$ of
Eq.~(\ref{Ured}) in case of the unbounded perfect Markovian spectral
coupling is modified into the contour $C_1$. The poles are denoted
schematically by the stars.} \label{fig3}
\end{figure}
An immediate consequence of this fact is that the modified
integration path $C_1$ of the background term $\int_{C_1} R^{\rm
II}(z){\rm e}^{-izt} dz$ in Eq.~(\ref{UredWW}) may be taken {\it
parallel} to the real axis ${\rm Im}~z=0$, and if there is no
obstacle to "dragging" it down to ${\rm Im}~z\rightarrow -\infty$,
the background contribution vanishes for all positive times.
Further, we focus on the contributions of the pole terms. Recall
Eq.~(\ref{R}) and note, that assuming this reduced equation is put
in the Markovian form, implies that the evolution generator $W(z)$
appearing in the denominator of Eq.~(\ref{R}) (similarly to the
total system propagator $U(z)={\cal L}_{t\rightarrow iz}{\rm
e}^{-iHt}=i/(z-H)$) is $z$-independent, even though not essentially
Hermitian. The $z$-independence of $W(z)$ automatically implies the
$z$-independence of its eigenvectors and eigenvalues defined in
Eq.~(\ref{eignW}), and hence, guarantees the orthogonality of the
projectors $Q_{\alpha_j}(z_j),~Q_{\alpha_k}(z_k)$ if
$\omega_{\alpha_j}\neq\omega_{\alpha_k}$ for any $z_j,z_k$.
Therefore, the idealized Markovian limit allows pure semigroup
evolution with no Zeno effect \cite{Adler} independently of the
position or the number
of the poles.\\

To prove our last conclusion explicitly we establish the connection
between the reduced generator $W^{\rm II}(z)$ and the spectral
correlation matrix in the lower half of the Laplace plane. Combining
Eq.~(\ref{R2},\ref{Rvsh2},\ref{h2}) we find
\begin{equation}
W^{\rm II}(z)=H_0^{red}+i\alpha(z).\label{Wvsalpha}
\end{equation}
Generally, the spectral correlation matrix $\alpha(t)$
(Eq.~(\ref{alphaI})), standing for the memory kernel of the
integro-differential equation (\ref{DotR}), may be any spread
function of time. The essence of the Markovian limit consists of
collapsing (either exactly or approximately) this memory-kernel into
a delta-correlated matrix. For the ``artificially" unbounded
spectrum $\lambda\in(-\infty,\infty)$ we have
\begin{equation}
\alpha^{\rm
Markov}(t)\int_{-\infty}^{\infty}d\lambda\omega(\lambda){\rm
e}^{-i\lambda t}=2\pi \omega\delta(t),\label{Mpoalpha}
\end{equation}
and hence,
\begin{equation}
i\alpha^{Markov}(z)=i2\pi\omega\int_{-\infty}^0 dt \delta(t) {\rm
e}^{izt}-2\pi i\omega=-\pi i\omega\label{Malphaz2}.
\end{equation}
where the spectral density matrix $\omega$ is a constant.
Substituting this back into Eq.~(\ref{Wvsalpha}) yields
\begin{equation}
W^{Markov{\rm
II}}(z)=H_0^{red}+i\alpha^{Markov}(z)=const.\\\label{MWvsalpha}
\end{equation}

What is left is to return to the original formulation of the
Wigner-Weisskopf theory with the semi-bounded continuous spectrum
and explain how the effect of weak coupling and resonance explain
the high precision of semigroup evolution either for the single or
for the many channel decay. In reaching the above result, it was
assumed $\omega(\lambda)$ is supported in $(-\infty,\infty)$. We
prove in Appendix A that $h^{\rm II}(z)$ may have zero determinant
for some values of $z$ in the lower half plane. The leading behavior
of the reduced evolution is therefore dictated only by a finite
number of the small regions corresponding to the poles of $R^{\rm
II}(z)$, where the contribution of the spectral density matrix
$\omega(\lambda)$ is strongly enhanced, in agreement with
resonance-Markovian approximation. In such a case, the integral over
$\lambda$ in Eq.~(\ref{alphaz2}) contributes primarily on the set of
$\lambda$'s close to the real part of the poles, whereas a
fictitious extension of the integration to $-\infty$ will not change
appreciably the value of the integral. Hence, it is straightforward
to repeat the derivations leading to the conclusion on the
simultaneous vanishing of the background contribution and the
orthogonality of the pole term projectors. The only difference is
that instead of the constant spectral density matrix $\omega$ in
Eqs.~(\ref{Mpoalpha},\ref{Malphaz2}) we shall have the resonant
$\lambda$-independent matrix $\omega^{Res}$ given by
Eq.~(\ref{omegaR}). To conclude, the result Eq.~(\ref{MWvsalpha}) is
quite general in case the resonances are fairly sharp (i.e., the
poles in the second sheet are close to the real line). Even though
not strictly exact, this Markovian limit approximation should be
valid for many experimental conditions, and
becomes even better for weaker coupling.\\

\section{Summary}

By associating the spectral weights of the Wigner-Weisskopf model
for unstable system decay with the statistical properties of
coupling to an environment, we were able to characterize the reduced
evolution in the subspace of unstable states in terms of the
spectral correlation matrix. Exploiting the properties of the
latter, we showed that the Markovian limit distribution is
sufficient to account for semigroup behavior for an arbitrary number
of the decay channels, observed in experiment, such as in
\cite{Fermilab}.
%It shall be noted, that there exists a rigorous underground for such
%interpretation of the Lee-Friedrichs model, which is called the
%dilation procedure. In case the reduced evolution is a non-unitary
%semigroup (i.e. Markovian), then the system may be imbedded into a
%higher-dimensional Hilbert where the evolution is unitary by the
%so-called Kolmogorov Dilation technique, which was applied in
%\cite{Maasen} for a damped Harmonic Oscillator. For non-Markovian
%reduced evolution the Kolmogorov dilation is inappropriate and the
%more general dilation methods are
%required.\\
Besides the important impact of the spectral correlation matrix
notion for understanding the reduced evolution of the multi-channel
decays, such as the two-channel K-meson decay \cite{CH,MH}, it is
clear that the association of the coupling to a spectral continuum
with an environment may be similarly useful for any analogously
modeled theory.% where any kind of projection on a subspace is
%invoked,
%regardless of the physical interpretation of the Hilbert space.\\
% As an interesting
%example we can point on the utilized Lax-Phillips type semigroups
%for the description of time evolution of resonances cite{Yossi,YLE}
%(and the references therein).

\section{Acknowledgements}
We thank Dr. Y. Strauss for discussion of analogous results obtained
by Nagy-Foias-Kolmogorov dilation \cite{R3}, in which the quantum
mechanical problem is imbedded in a larger space as in the work of
Maassen \cite{Maasen}, who constructs a Fock space resulting in the
generation of a ``noisy environment" of the type we
have discussed.% by a bath modeled by coherent states of a Fock
%space.

\section{Appendix A. Second sheet properties of $\textsc{\textbf{R(z)}}$}
\setcounter{equation}{0}
\renewcommand{\theequation}{A.\arabic{equation}}

In this appendix we discuss the structure of the reduced propagator
in the complex Laplace plane. We specifically show that the
propagator has poles only in the lower half plane near the real axis
and also provide the background needed for establishing the
connection between the reduced system evolution generator and the
spectral density matrix Eq.~(\ref{Wvsalpha}). Following the
well-known procedure \cite{SH} let us rewrite Eq.~(\ref{R}) as
\begin{equation}
R(z)=\frac{1}{h(z)}.\label{Rvsh}
\end{equation}
Here the operator
\begin{equation}
h(z)=z-H_{0}^{red}- \int{\rm
d}\lambda\frac{\omega(\lambda)}{z-\lambda},\label{h}
\end{equation}
with $H_{0}^{red}$ and $\omega(\lambda)$ given by
Eqs.~(\ref{H0S},\ref{hatomega}) respectively, is found
straightforwardly from the projection of the total system propagator
on the discrete subspace. The last term of Eq.~(\ref{h}) can be
recognized as the Laplace transform (for ${\rm Im}z>0$) of the
spectral correlation matrix $\alpha(t)$ Eq.~(\ref{alphaI}):
$$
i\alpha(z)\equiv \int d\lambda\frac{\omega(\lambda)}{z-\lambda}=i
\int_0^\infty dt \alpha(t) e^{izt}=$$\begin{equation}=i\int_0^\infty
dt \int{\rm d}\lambda\omega(\lambda) {\rm e}^{-i\lambda t}{\rm
e}^{izt}\label{alphaz}.
\end{equation}
To obtain a singularity in $R(z)$ Eq.~(\ref{Rvsh}) we need that the
determinant of $h(z)$ Eq.~(\ref{h}) vanishes. These non-Hermitian
matrices can be put into Jordan canonical form with a unitary
transformation, with eigenvalues along the diagonal. A vanishing
eigenvalue implies a vanishing determinant. We therefore can ask
whether $h(z)$ has a vanishing eigenvalue, i.e.
\begin{equation}
h(z)|\chi(z)\rangle_{\rm R}=\left[z-H_{0}^{red}- \int{\rm
d}\lambda\frac{\omega(\lambda)}{z-\lambda}\right]|\chi(z)\rangle_{\rm
R}=0,\label{hchi}
\end{equation}
where $|\chi(z)\rangle_{\rm R}$ is the right eigenvector of $h(z)$
with the presumed zero eigenvalue. Taking the scalar product with
$_{\rm L}\langle\chi(z)|$ eigenvector from the left yields
$$
z|\chi(z)\rangle|^2-_{\rm
L}\langle\chi(z)|H_{0}^{red}|\chi(z)\rangle_{\rm
R}-$$\begin{equation}- \int{\rm d}\lambda\frac{_{\rm
L}\langle\chi(z)|\omega(\lambda)|\chi(z)\rangle_{R}}{z-\lambda}=0.\label{chihchi}
\end{equation}
In general, for the convergence of the integral in the last
expression, we must assume that $\omega(\lambda)$ decreases better
than an, no matter how small, inverse power of $\lambda$ since the
integral contains $\sim\frac{d\lambda}{\lambda}$ for very large
$\lambda$. Now we consider the imaginary part of
Eq.~(\ref{chihchi}). Since $H_{0}^{red}$ is Hermitian, its
expectation value in the state $|\chi(z)\rangle_{\rm R}$ is real,
and we are left with
\begin{equation}
{\rm Im}z\left\{|\chi(z)\rangle|^2+ \int{\rm d}\lambda\frac{_{\rm
L}\langle\chi(z)|\omega(\lambda)|\chi(z)\rangle_{R}}{|z-\lambda|^2}\right\}=0,\label{Imchihchi}
\end{equation}
i.e., ${\rm Im}~z$ times a positive quantity. This can never be zero
for ${\rm Im}~z\neq0$ on the first Riemann sheet. We are therefore
required to go to the second sheet:
\begin{equation}
R^{\rm II}(z)=\frac{1}{h^{\rm II}(z)},\label{Rvsh2}
\end{equation}
\begin{equation}
h^{\rm II}(z)=z-H_{0}^{red}- i\alpha(z).\label{h2}
\end{equation}
Here $R^{\rm II}(z)$ and $h(z)^{\rm II}$ are the smooth
continuations of $R(z)$ and $h(z)$ into the lower half plane and
\begin{equation}
i\alpha(z)=i\int_{-\infty}^0 dt \int d\lambda\omega(\lambda) {\rm
e}^{-i\lambda t}{\rm e}^{izt}-2\pi i\omega(z)\label{alphaz2}.
\end{equation}
is the analytic continuation of the spectral correlation matrix
$\alpha(z)$ Eq.~(\ref{alphaz}), where $\omega(z)$ is the analytic
continuation of $\omega(\lambda)$ into the lower half plane. The
condition for a vanishing eigenvalue Eq.~(\ref{chihchi}) now reads
$$
z|\chi(z)\rangle|^2-_{\rm
L}\langle\chi(z)|H_{0}^{red}|\chi(z)\rangle_{\rm
R}-$$\begin{equation}- \int{\rm d}\lambda\frac{_{\rm
L}\langle\chi(z)|\omega(\lambda)|\chi(z)\rangle_{R}}{z-\lambda}+
2\pi i _{\rm
L}\langle\chi(z)|\omega(z)|\chi(z)\rangle_{R}=0,\label{chihchi2}
\end{equation}
Taking the imaginary part as before gives
$$
{\rm Im}z\left\{|\chi(z)\rangle|^2+ \int{\rm d}\lambda\frac{_{\rm
L}\langle\chi(z)|\omega(\lambda)|\chi(z)\rangle_{R}}{|z-\lambda|^2}\right\}+$$\begin{equation}+2\pi
i _{\rm
L}\langle\chi(z)|\omega(z)|\chi(z)\rangle_{R}=0,\label{Imchihchi}
\end{equation}
so that there may be a solution for ${\rm Im}z$ negative, since
$\omega(\lambda)$ is positive definite. It is easy to see that the
expectation of $\omega(\lambda)$ is the sum of absolute squares; the
analytic continuation of $\omega(\lambda)$ to the lower half plane
for small imaginary part of $z$, enough to reach a resonance pole,
is assumed to remain approximately real since it is smoothly
connected to real values of $\omega(\lambda)$, for $\lambda$ on the
real line. In the Markovian limit this function is taken to
approach a constant.\\

\end{document}